\documentclass[journal]{IEEEtran}
\IEEEoverridecommandlockouts

\usepackage{graphicx}
\usepackage{graphbox}
\usepackage[caption=false,font=footnotesize, subrefformat=parens]{subfig}
\usepackage{cite}
\usepackage{booktabs}
\usepackage{url}
\usepackage{amsmath}
\usepackage{amssymb}
\usepackage{physics}

\newcommand{\scr}[1]{\mathrm{#1}}       
\newcommand{\pl}{\mathrm{+}}
\newcommand{\mi}{\mathrm{-}}

\DeclareMathOperator{\RE}{Re}
\DeclareMathOperator{\IM}{Im}
\DeclareMathOperator{\SIGN}{sign}
\newcommand{\D}{\scr{d}}
\newcommand{\jj}{\scr{j}}
\newcommand{\e}{\scr{e}}                
\newcommand{\kc}{k_{\scr{c}}}
\newcommand{\ko}{\boldsymbol{k}_{\scr{o}}}
\newcommand{\kpb}{\boldsymbol{k}_{\scr{p}}}
\newcommand{\kpbo}{\boldsymbol{k}_{\scr{p0}}}
\newcommand{\kp}{k_{\scr{p}}}
\newcommand{\kvb}{\boldsymbol{k}_{\scr{v}}}
\newcommand{\kvbo}{\boldsymbol{k}_{\scr{v0}}}
\newcommand{\kv}{k_{\scr{v}}}
\newcommand{\alphac}{\alpha_\mathrm{c}}
\newcommand{\alphap}{\alpha_\mathrm{p}}
\newcommand{\betapv}{\beta_\mathrm{pv}}
\newcommand{\betavp}{\beta_\mathrm{vp}}
\newcommand{\alphav}{\alpha_\mathrm{v}}

\newcommand{\omegao}{\omega_\scr{o}}
\newcommand{\omegag}{\omega_\scr{g}}
\newcommand{\omegac}{\omega_\scr{c}}
\newcommand{\thetac}{\vartheta_\scr{c}}

\newcommand{\icref}{\boldsymbol{i}_\mathrm{c,ref}}
\newcommand{\icreflim}{\bar{\boldsymbol{i}}_\mathrm{c,ref}}
\newcommand{\Ts}{{T_\mathrm{s}}}

\newcommand{\ug}{\boldsymbol{u}_\mathrm{g}}
\newcommand{\ugp}{\boldsymbol{u}_\mathrm{g}^\pl}
\newcommand{\ugn}{\boldsymbol{u}_\mathrm{g}^\mi}
\newcommand{\ugpo}{\boldsymbol{u}_\mathrm{g0}^\pl}
\newcommand{\hatugp}{\hat{\boldsymbol{u}}_\mathrm{g}^\pl}
\newcommand{\hatugn}{\hat{\boldsymbol{u}}_\mathrm{g}^\mi}
\newcommand{\tildeugp}{\tilde{\boldsymbol{u}}_\mathrm{g}^\pl}
\newcommand{\tildeugpo}{\tilde{\boldsymbol{u}}_\mathrm{g0}^\pl}
\newcommand{\tildeugn}{\tilde{\boldsymbol{u}}_\mathrm{g}^\mi}
\newcommand{\tildeugno}{\tilde{\boldsymbol{u}}_\mathrm{g0}^\mi}
\newcommand{\hatug}{\hat{\boldsymbol{u}}_\mathrm{g}}

\newcommand{\ugmago}{u_\mathrm{g0}}
\newcommand{\ugmag}{u_\mathrm{g}}

\newcommand{\ic}{\boldsymbol{i}_\mathrm{c}}
\newcommand{\icp}{\boldsymbol{i}_\mathrm{c}^\pl}
\newcommand{\icpo}{\boldsymbol{i}_\mathrm{c0}^\pl}
\newcommand{\icn}{\boldsymbol{i}_\mathrm{c}^\mi}
\newcommand{\icno}{\boldsymbol{i}_\mathrm{c0}^\mi}

\newcommand{\vcn}{\boldsymbol{v}_\mathrm{c}^\mi}
\newcommand{\vcno}{\boldsymbol{v}_\mathrm{c0}^\mi}
\newcommand{\hatvcn}{\hat{\boldsymbol{v}}_\mathrm{c}^\mi}
\newcommand{\vcp}{\boldsymbol{v}_\mathrm{c}^\pl}
\newcommand{\vcpo}{\boldsymbol{v}_\mathrm{c0}^\pl}
\newcommand{\vcpmag}{v_\mathrm{c}^\pl}
\newcommand{\vcpmago}{v_\mathrm{c0}^\pl}
\newcommand{\hatvcp}{\hat{\boldsymbol{v}}_\mathrm{c}^\pl}
\newcommand{\hvcpmag}{\hat{v}_\mathrm{c}^\pl}
\newcommand{\hvcpmago}{\hat{v}_\mathrm{c0}^\pl}

\newcommand{\vc}{\boldsymbol{v}_\mathrm{c}}
\newcommand{\vco}{\boldsymbol{v}_\mathrm{c0}}
\newcommand{\vcmago}{v_\mathrm{c0}}
\newcommand{\hatvc}{\hat{\boldsymbol{v}}_\mathrm{c}}
\newcommand{\uc}{\boldsymbol{u}_\mathrm{c}}
\newcommand{\vcref}{v_\mathrm{c,ref}}
\newcommand{\ucref}{\boldsymbol{u}_\mathrm{c,ref}}
\newcommand{\ucrefp}{\boldsymbol{u}_\mathrm{c,ref}^\pl}
\newcommand{\ucrefn}{\boldsymbol{u}_\mathrm{c,ref}^\mi}
\newcommand{\ucsref}{\boldsymbol{u}_\mathrm{c,ref}^\mathrm{s}}
\newcommand{\pgref}{p_\mathrm{g,ref}}
\newcommand{\pgreflim}{\bar{p}_\mathrm{g,ref}}
\newcommand{\pg}{p_\mathrm{g}}
\newcommand{\qg}{q_\mathrm{g}}
\newcommand{\pgp}{p_\mathrm{g}^\pl}
\newcommand{\pgpo}{p_\mathrm{g0}^\pl}
\newcommand{\hatpgp}{\hat{p}_\mathrm{g}^\pl}
\newcommand{\hatpgpo}{\hat{p}_\mathrm{g0}^\pl}

\newcommand{\ifilt}{\boldsymbol{i}_\mathrm{f}}
\newcommand{\Lg}{L_\mathrm{g}}
\newcommand{\hL}{\hat{L}}
\newcommand{\tL}{\tilde{L}}

\newcommand{\ec}{\boldsymbol{e}_\mathrm{c}}

\newcommand{\zetao}{\zeta_\mathrm{o}}

\newcommand{\eo}{\boldsymbol{e}_\scr{o}}

\newcommand{\knb}{\boldsymbol{k}_{\scr{n}}}
\newcommand{\kn}{k_{\scr{n}}}
\newcommand{\alphan}{\boldsymbol{\alpha}_\mathrm{n}}

\newcommand{\betapn}{\boldsymbol{\beta}_\mathrm{pn}}
\newcommand{\betavn}{\boldsymbol{\beta}_\mathrm{vn}}
\newcommand{\betanp}{\boldsymbol{\beta}_\mathrm{np}}
\newcommand{\betanv}{\boldsymbol{\beta}_\mathrm{nv}}

\newcommand{\knbo}{\boldsymbol{k}_{\scr{n}}}
\newcommand{\ucreflim}{\bar{u}_\mathrm{c,ref}}

\newcommand{\rkno}{\boldsymbol{k}_\mathrm{n}}
\newcommand{\rknoe}{\boldsymbol{k}_\mathrm{n}'}
\newcommand{\ru}{\Delta\boldsymbol{u}}
\newcommand{\rvcpo}{v_\mathrm{c0}^\pl}
\newcommand{\rugdo}{u_\mathrm{gd0}^\pl}
\newcommand{\af}{\boldsymbol{a}_\mathrm{f}}

\begin{document}

\title{Disturbance-Observer-Based Grid-Forming Control for Unbalanced Grids}

\author{
	\vskip 1em
	
  Juho~Määttä, \emph{Graduate Student Member, IEEE}, Marko~Hinkkanen, \emph{Fellow, IEEE}, Tuure~Nurminen, \\
  Orcun~Karaca, \emph{Senior Member, IEEE}, Rayane~Mourouvin, \emph{Member, IEEE}, \\
  Jarno~Kukkola, and~Lennart Harnefors, \emph{Fellow, IEEE}

  \thanks{This work has been submitted to the IEEE for possible publication. Copyright may be transferred without notice, after which this version may no longer be accessible.
  
  This project was supported by ABB Oy. The authors acknowledge the use of the EPE infrastructure of Aalto University School of Electrical Engineering.

  J. Määttä and M. Hinkkanen are with the Department of Electrical Engineering and Automation, Aalto University School of Electrical Engineering, 02150 Espoo, Finland (e-mail: juho.k.maatta@aalto.fi; marko.hinkkanen@aalto.fi).

  T. Nurminen and J. Kukkola are with ABB Oy Drives, 00380 Helsinki, Finland (e-mail: tuure.nurminen@fi.abb.com; jarno.kukkola1@fi.abb.com).

  O. Karaca is with ABB Corporate Research, 5405 Baden, Switzerland (e-mail: orcun.karaca@ch.abb.com).

  R. Mourouvin is with Aix Marseille Univ, CNRS, LIS, Marseille, France (e-mail: rayane.mourouvin@univ-amu.fr).

  L. Harnefors is with ABB Corporate Research, 72226 Västerås, Sweden (e-mail: lennart.harnefors@se.abb.com).
  }
}

\maketitle

\begin{abstract}
This article proposes a grid-forming control method for operation under unbalanced grid-voltage conditions. The method regulates the positive-sequence active power delivered to the grid and actively suppresses the negative-sequence converter voltage, thereby maintaining a constant converter-voltage magnitude during unbalanced faults when the converter operates within its physical limits. Only the converter current is measured on the AC side, and a disturbance observer is used to synchronize with the grid and to provide integral and resonant action. Estimates for the positive- and negative-sequence grid voltage are obtained from the disturbance observer. A current-limitation scheme for both balanced and unbalanced faults is integrated. Comprehensive stability analysis and tuning guidelines are provided. Experimental results using a 12.5-kVA converter demonstrate that the proposed method can operate during severe balanced and unbalanced faults.
\end{abstract}


\section{Introduction}

\IEEEPARstart{T}{he increase} in the share of converter-interfaced generation has pushed grid operators worldwide to publish specifications for grid-forming converters \cite{AEMO2023,GBgrid1,fingrid-sjv2024}. A common requirement for grid-forming converters is operation as a voltage source behind an impedance, and ideally the converter should support the grid by maintaining a purely positive-sequence voltage even during unbalanced grid-voltage conditions \cite{fingrid2026,GBgrid2,Unifi2026}. This requires the grid-forming converter to allow negative-sequence current to flow \cite{fingrid2026,Unifi2026,entsoe-gfm2025}.

Control methods that consider unbalanced grid-voltage conditions often require that measured quantities can be decomposed into their positive- and negative-sequence components. 
This can be achieved with sequence-extraction methods, which were originally introduced for grid-following control methods to extend the classical synchronous-reference-frame phase-locked loop.
One example of such a method is the decoupled double-synchronous-reference-frame phase-locked loop \cite{Rodriguez2007}, which uses separate reference frames to obtain the positive- and negative-sequence components. 
Sequence extraction can also be achieved using band-pass filters, and examples of such methods are the second-order generalized integrator \cite{Rodriguez2011} and the multiple complex-coefficient filter phase-locked loop \cite{Guo2011}. A review of sequence-extraction methods can be found in \cite{Xin2016_2} and a comparative study in \cite{Mel2022}.

Examples of grid-forming control methods utilizing such sequence-extraction methods can be found in \cite{Rosso2021,Qoria2023,Awal2024,Khan2025,Zhang2026,Liu2026,Gao2026,Ojo2026}. These methods require measurement of the voltage on the AC side of the converter. This measurement may not be available, for example, because of hardware constraints and cost-saving considerations. Moreover, as the number of measured signals increases, so does the complexity of the control system since a minimum of two real-valued states (delays or integrators) is required to separate a complex-valued quantity into positive and negative sequences.

An alternative to extracting sequences from the measured voltage is to use a disturbance observer to estimate the positive- and negative-sequence components of an input-equivalent disturbance.  
The methods proposed in \cite{Rios2023,Rios2024,Nurminen2025} are recent examples of grid-forming control methods using such observer-based sequence-extraction methods.
The method in \cite{Rios2023} uses dual-loop cascade control with outer voltage-control and inner current-control loops, with the disturbance observer implemented as a Kalman filter.
The method in \cite{Rios2024} uses single-loop voltage control and switches to a current-control mode during faults. The grid-forming and current-control modes utilize separate observers, both implemented as Kalman filters. Switching between the two control modes is implemented with a fault-detection algorithm, which increases tuning complexity and can introduce chattering during mode switching.
Both \cite{Rios2023} and \cite{Rios2024} demonstrate a wide operating range in both strong and weak grids, as well as during balanced and unbalanced grid faults. However, since both methods model the LC filter of the converter, the control systems are relatively high-order with 10 and 20 (real-valued) state variables, respectively. These methods also use full-state measurement, measuring both the capacitor voltage and grid current in addition to the converter current.

In \cite{Nurminen2025}, the disturbance observer is shared between grid-forming and current-control modes. The method assumes a simpler L filter and requires only the converter current measurement, but lacks dynamic sequence separation of the converter current. When negative-sequence grid voltage is present, the method can maintain sinusoidal output currents only in grid-forming mode and not in current-control mode. Current limitation in the periodic steady state is realized by limiting the active-power and voltage references. The control method, including the controlled outputs, the control law, and the reference limitation, is designed based on the periodic steady-state solution and the assumption that the converter voltage contains no negative-sequence component. Therefore, the method is not well defined during transients. Furthermore, comprehensive stability analysis of the method is not included.

The overcurrent capability of converters is inherently limited by the physical constraints of semiconductor devices. Grid faults often cause large currents that must thus be limited to protect the converter hardware. Three common current-limitation methods for grid-forming converters are a current-control mode \cite{Rosso2021,Avdiaj2022,Qoria2023,Rios2024,Stanojev2025,Ojo2026}, virtual impedance \cite{Zarei2019,Zubi2021,Morales2024,Khan2025,Zhang2026}, and limitation of the references provided to the grid-forming control method \cite{Chen2020,Taul2020_1,Rios2023,Nar2023,He2025,Liu2026,Gao2026,Qoria2023}.
A review of current-limitation methods for grid-forming converters can be found in \cite{Baeckeland2024}.
Limiting the references has the advantage of inherently improving transient stability by providing realizable references \cite{Huang2019,Rokrok2022,Qoria2023}; this approach is adopted in the proposed control scheme.

\begin{figure}[!t]
    \centering
    \includegraphics{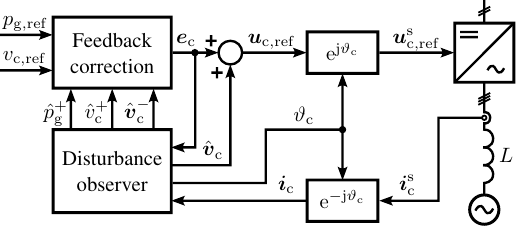}
    \caption{Proposed control method in grid-forming control mode.} \label{fig:ctrl}
\end{figure}

None of the methods in \cite{Rosso2021,Qoria2023,Rios2023,Rios2024,Awal2024,Khan2025,Zhang2026,Liu2026,Gao2026,Ojo2026,Nurminen2025} combines the following features: only the converter current is measured, the same low-order control states are used both in normal operation and during current limitation, and the negative-sequence converter voltage is suppressed using dynamic sequence separation.
To address these missing features, this article proposes a grid-forming control method for unbalanced grid conditions. The 
method extends the approach of \cite{Nur2024} from balanced to unbalanced grids. The control structure is shown in Fig.~\ref{fig:ctrl}. The main contributions are as follows:

\begin{itemize}
    \item The proposed grid-forming control method uses, unlike \cite{Rosso2021,Qoria2023,Rios2023,Rios2024,Awal2024,Khan2025,Zhang2026,Liu2026,Gao2026,Ojo2026}, a disturbance observer to unify synchronization and estimation of both the positive and negative sequences of the grid voltage. Conventional phase-locked loops and AC-side voltage measurement are thereby avoided. In contrast to \cite{Rios2024}, the disturbance observer, and thus all control states, are shared between the grid-forming and current-control modes. Therefore, having different states for different control modes is avoided, and the order of the control system is reduced so that it requires only eight (real-valued) state variables.
    \item In contrast to \cite{Rios2023}, a single-loop controller is used to directly control the positive-sequence voltage magnitude and active power. It actively suppresses the negative-sequence converter voltage, allowing the converter to maintain positive-sequence voltage-source characteristics under unbalanced grid conditions, within the physical limits of the converter.
    \item The method includes a current-limitation strategy for both balanced and unbalanced faults. Unlike \cite{Nurminen2025}, where the current limitation is based on the periodic steady-state solution requiring additional assumptions, dynamic sequence separation is employed to obtain positive- and negative-sequence current references, allowing the sequence currents to be limited independently while maintaining sinusoidal currents. Active-power reference limitation is employed to provide a realizable reference during faults, enhancing transient stability \cite{Huang2019,Rokrok2022,Qoria2023}.
    \item Stability analysis and tuning guidelines are presented, yielding explicit stability conditions for the feasible operating region. The analysis also shows that the method tolerates severe under- and overestimation of the inductance.
\end{itemize}

Experimental results on a 12.5-kVA converter demonstrate the operation of the method. The results verify the performance under balanced and unbalanced faults in both strong and very weak grids, using the same inductance estimate. Simulation models for the same test cases are openly available as a Code Ocean capsule \cite{motulator_capsule}.

The remainder of this article is organized as follows. Section~\ref{chap:grid_model} presents the system model and the dynamic sequence separation. Section~\ref{chap:control} details the proposed disturbance-observer-based control method. Section~\ref{chap:lin} provides the small-signal stability analysis. Implementation aspects, including the reference-limitation strategy, are discussed in Section~\ref{chap:implementation}. Section~\ref{chap:results} presents the experimental results, and Section~\ref{chap:conclusion} concludes the article.

\begin{figure}[t]
    \centering
    \includegraphics{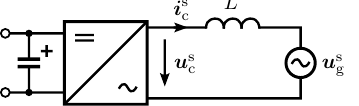}
    \caption{System model in stationary coordinates, indicated by superscript s.}
    \label{fig:grid}
\end{figure}

\section{System Model} \label{chap:grid_model}

\subsection{State and Output Equations}
Complex space vectors are used and per-unit quantities are assumed. Fig.~\ref{fig:grid} shows the system model, consisting of a voltage-source converter connected to an inductive grid. The total inductance $L$ consists of the filter and grid inductances. In the following, this system is modeled in general coordinates, whose angle is $\thetac$ with respect to the stationary coordinates and which rotate at the angular speed $\omegac = \D\thetac/\D t$.  

The dynamics of the converter current are governed by  
\begin{subequations} \label{eq:model}
\begin{align} \label{eq:grid_ic}
    L\frac{\D\ic}{\D t} &= \uc - \ug - \jj\omegac L\ic 
\end{align}
where $\ic$ is the converter current and $\uc$ is the converter voltage. The grid voltage $\ug$ is assumed to consist of both positive-sequence voltage $\ugp$ and negative-sequence voltage $\ugn$. Correspondingly, the grid-voltage dynamics are
\begin{align} \label{eq:grid_ug}
    \frac{\D\ugp}{\D t} &= \jj(\omegag - \omegac)\ugp \\
    \frac{\D\ugn}{\D t} &= -\jj(\omegag + \omegac)\ugn \\
    \ug &= \ugp + \ugn
\end{align}
\end{subequations}
where $\omegag$ is the grid angular frequency. With the adopted per-unit scaling, the instantaneous active power delivered to the grid is $\pg=\Re\{\ug\ic^*\}$. In unbalanced conditions, $\pg$ may contain oscillating components.

\subsection{Augmentation With Dynamic Sequence Separation} 
The instantaneous converter current $\ic$ depends on the model \eqref{eq:model}, with input $\uc$ and disturbance $\ug$. In the periodic steady state, omitting the switching harmonics, the converter current (and other space vectors) can be separated into sequence components, i.e., $\ic = \icp + \icn$. The positive- and negative-sequence components have constant magnitudes and rotate at the angular frequencies $\omegag-\omegac$ and $-\omegag-\omegac$, respectively.

For control synthesis, it is useful to define quasi-static sequence currents, which are also available during transients. First, a complex low-pass filter for the converter current is defined as
\begin{subequations} \label{eq:augmented_model}
\begin{align} \label{eq:sequence_filter}
    \frac{\D \ifilt}{\D t} = -(\omegag + \jj\omegac)\ifilt + \omegag \ic
\end{align}
where $\ifilt$ is the filtered current and the filter bandwidth equals the grid angular frequency $\omegag$. Using this filtered current, the quasi-static positive-sequence and negative-sequence currents are defined as~\cite{Golestan2020}
\begin{align} \label{eq:icp_def}
    \icp &= \frac{\ic}{1 + \jj} + \jj\ifilt \\
    \icn &= \frac{\ic}{1 - \jj} - \jj\ifilt. \label{eq:icn_def}
\end{align}
These are dynamic output quantities that in the periodic steady state converge to the actual positive- and negative-sequence currents. The identity $\ic = \icp + \icn$ also holds during transients. 

Using the dynamic sequence separation \eqref{eq:augmented_model}, other quasi-static quantities can be defined. From the grid-forming control perspective, the quasi-static converter voltage $\vc$ and its sequence components $\vcp$ and $\vcn$ are of particular interest: 
\begin{align} 
    \vcp &= \ugp + \jj\omegag L\icp \label{eq:vcp} \\
    \vcn &= \ugn - \jj\omegag L\icn \label{eq:vcn} \\
    \vc &= \vcp + \vcn = \ug + \jj\omegag L(\icp - \icn).
\end{align}
Furthermore, the positive-sequence active power is 
\begin{align} \label{eq:pg_def}
    \pgp &= \Re{\ugp(\icp)^*}. 
\end{align}
\end{subequations}
The quantities defined in \eqref{eq:icp_def}--\eqref{eq:pg_def} are used as controlled outputs of the system consisting of \eqref{eq:model} and \eqref{eq:augmented_model}. These outputs also have a clear physical interpretation in the periodic steady state.

\subsection{Dynamics of Selected Outputs}
For the purposes of control synthesis and analysis, the dynamics of the selected quasi-static outputs are derived using \eqref{eq:model} and \eqref{eq:augmented_model}. The dynamics of the positive-sequence active power $\pgp$ and voltage magnitude $\vcpmag = |\vcp|$ are
\begin{subequations} \label{eq:output_dyn}
\begin{align} \label{eq:pg_dyn}
    \frac{\D\pgp}{\D t} &= \frac{1}{L}\Re{\frac{(\ugp)^*}{1 + \jj}(\uc - \vc)} \\ 
    \frac{\D\vcpmag}{\D t} &= \frac{\omegag}{\vcpmag}\Re{\frac{(\vcp)^*}{1-\jj}(\uc - \vc)} . \label{eq:vcp_dyn}  
\end{align}
Furthermore, the negative-sequence converter-voltage vector has the dynamics
\begin{align}
    \frac{\D\vcn}{\D t} &= \frac{\omegag}{1 + \jj}(\uc - \vc) - \jj(\omegag + \omegac)\vcn . \label{eq:vcn_vec_dyn}
\end{align}
\end{subequations}
These equations are convenient when studying the closed-loop dynamics of the control system, as they explicitly show how the control input $\uc$ affects the selected outputs. If other outputs are of interest, their dynamics can be derived in a similar manner. 

\section{Proposed Control Method} \label{chap:control}
This section presents the proposed control system. Two control modes are considered: a grid-forming mode, shown in Fig.~\ref{fig:ctrl}, for regulating positive-sequence active power and converter-voltage magnitude, and a current-control mode for regulating the converter output current. The disturbance observer is shared between the control modes.

\subsection{Disturbance Observer}
A disturbance observer is used to estimate the positive- and negative-sequence components of the grid voltage, to synchronize with the grid, and to provide both integral and resonant action to the proposed control method. The disturbance observer can be formulated based on~\eqref{eq:model} as
\begin{subequations} \label{eq:obs_cont}
\begin{align}
    \frac{\D \hatugp}{\D t} &= \jj(\omegag - \omegac)\hatugp + \boldsymbol{k}_\mathrm{o1} \eo \\ 
    \frac{\D \hatugn}{\D t} &= -\jj(\omegag + \omegac)\hatugn + \boldsymbol{k}_\mathrm{o2} \eo \\
    \eo &= \ucref - \hatug - \hat L \frac{\D\ic}{\D t} - \jj\omegac \hat L \ic \label{eq:obs_error}
\end{align}
where $\boldsymbol{k}_\mathrm{o1}$ and $\boldsymbol{k}_\mathrm{o2}$ are the observer gains. Throughout the article, estimated quantities are marked with hats. Using~\eqref{eq:augmented_model}, the quasi-static voltage estimates are
\begin{align}
    \hatvcp &= \hatugp + \jj\omegag \hat L \icp \\ 
    \hatvcn &= \hatugn - \jj\omegag \hat L \icn \\
    \hatvc &= \hatvcp + \hatvcn.
\end{align}
The positive-sequence active-power estimate is 
\begin{align}
    \hatpgp &= \Re{\hatugp(\icp)^*}
\end{align}
\end{subequations}
and the positive-sequence voltage-magnitude estimate is $\hvcpmag = |\hatvcp|$. 

In the implementation, the direct discrete-time variant of this observer derived in Appendix~\ref{app:zoh} is used. This improves the performance at low sampling frequencies, compared with the continuous-time variant adopted in \cite{Nurminen2025}.

\subsection{Grid-Forming Control}
The control target in grid-forming mode is to regulate the positive-sequence active power and the positive-sequence converter-voltage magnitude. The grid-forming control law is
\begin{subequations} \label{eq:gfm_control_law}
\begin{align}
    \ucref &= \hatvc + \ec^\pl + \ec^\mi \\
    \ec^\pl &= \kpb(\pgref - \hatpgp) + \kvb(\vcref - \hvcpmag) \\ 
    \ec^\mi &= -\knb\hatvcn
\end{align}
\end{subequations}
where $\ec = \ec^\pl + \ec^\mi$ is the feedback correction, $\pgref$ is the active-power reference, $\vcref$ is the voltage-magnitude reference, and $\knb = \jj\kn$ is the purely imaginary negative-sequence converter-voltage feedback gain with $\kn > 0$.

Based on the selected output dynamics \eqref{eq:output_dyn}, the term $\ec^\pl$ is used to regulate the positive-sequence active power and converter-voltage magnitude. The additional term $\ec^\mi$ provides feedback from the estimated negative-sequence converter voltage $\hatvcn$, which is zero when the converter voltage contains no negative-sequence component. Thus, $\ec^\mi$ can be interpreted as proportional feedback control of the negative-sequence converter voltage with zero reference.

As shown in Section~\ref{sec:dyn_linear}, the positive-sequence active-power and voltage-magnitude channels can be fully decoupled with the complex gains
\begin{align} \label{eq:gfm_gain_full}
    \kpb = \kp\frac{(1 + \jj)\hatvcp}{|\hatvcp|} \qquad \kvb = \kv\frac{(1-\jj)\hatugp}{|\hatugp|} .
\end{align}
Alternatively, partial decoupling can be used. The gain selection
\begin{align} \label{eq:gfm_gain_partial2}
    \kpb = \kp\frac{\hatvcp}{|\hatvcp|} \qquad
    \kvb = \kv\frac{(1-\jj)\hatugp}{|\hatugp|}
\end{align}
decouples the active-power-control channel from the voltage-control channel. The opposite partial decoupling is obtained with
\begin{align} \label{eq:gfm_gain_partial}
    \kpb = \kp\frac{(1+\jj)\hatvcp}{|\hatvcp|} \qquad \kvb = \kv\frac{\hatvcp}{|\hatvcp|}
\end{align}
which decouples the voltage-control channel from the active-power-control channel. The last alternative \eqref{eq:gfm_gain_partial} is used in the experiments. This selection provides robustness against inductance estimation errors, since the gains depend only on $\hatvcp$, which can be estimated exactly, as shown in Section~\ref{sec:inductance_error}.

\subsection{Transparent Current Control}
The current-control mode regulates the converter output current and is implemented as a transparent current controller that acts only if the converter current exceeds its maximum value. The current controller is defined as
\begin{subequations} \label{eq:CC_mode}
\begin{align} \label{eq:CC_mode_uc}
    \ucref = \hatvc + \kc\left(\icreflim^\pl - \icp + \icreflim^\mi - \icn\right)
\end{align}
where $\kc = \alphac\hL$ is the current-control gain, $\alphac$ is the current-control bandwidth, and $\icreflim^\pl$ and $\icreflim^\mi$ are the limited current references for the positive and negative sequences, respectively. The current-reference limitation is considered an implementation aspect and is thus detailed in Section~\ref{chap:implementation}. The unlimited current references are
\begin{align} \label{eq:icref}
    \icref^\pl = \icp + \frac{1}{\kc}\ec^\pl \qquad \icref^\mi = \icn + \frac{1}{\kc}\ec^\mi .
\end{align}
\end{subequations}
If the converter current stays below its allowed steady-state maximum value, the current-reference limiter passes the references through unchanged, i.e., $\icreflim^\pl = \icref^\pl$ and $\icreflim^\mi = \icref^\mi$ hold. Thus, the control law~\eqref{eq:gfm_control_law} is automatically restored and the converter operates in grid-forming mode.

\section{Stability Analysis} \label{chap:lin}
For analysis purposes, a linearized model for the closed-loop system consisting of \eqref{eq:model}, \eqref{eq:augmented_model}, \eqref{eq:obs_cont}, and \eqref{eq:gfm_control_law} is derived. The model is linearized around an operating point in the periodic steady state. Without loss of generality, coordinates are assumed to rotate at the grid angular frequency, $\omegac = \omegag$, in this section. Furthermore, accurate converter-voltage production $\ucref = \uc$ is assumed. A preceding $\Delta$ marks a small-signal deviation about the operating-point quantity, marked with a subscript $0$. Estimation errors are marked with a tilde. Sections~\ref{sec:est_dyn} and~\ref{sec:dyn_linear} assume that the inductance is known, i.e., $\hat L = L$. The effect of inductance estimation error is then studied in Section~\ref{sec:inductance_error}.

\subsection{Estimation-Error Dynamics} \label{sec:est_dyn}
Using the system model \eqref{eq:model} and the disturbance observer \eqref{eq:obs_cont}, the estimation-error dynamics of the positive- and negative-sequence grid voltage become
\begin{subequations} \label{eq:estimation_error_dyn}
\begin{align}
    \frac{\D \Delta\tildeugp}{\D t} &= -\boldsymbol{k}_\mathrm{o1}\left(\Delta\tildeugp+\Delta\tildeugn\right)\\
    \frac{\D \Delta\tildeugn}{\D t} &= -\boldsymbol{k}_\mathrm{o2}\Delta\tildeugp - \left(\boldsymbol{k}_\mathrm{o2} + \jj2\omegag\right)\Delta\tildeugn
\end{align}
\end{subequations}
where $\Delta\tildeugp = \Delta\ugp - \Delta \hatugp$ is the estimation error of the positive-sequence grid voltage and other estimation errors are defined similarly. There is no coupling from the control law to the observer estimation error. Consequently, the estimation error acts like an external disturbance from the viewpoint of the control system. 

The dynamics \eqref{eq:estimation_error_dyn} result in the characteristic equation 
\begin{align}
    s^2 + \left(\boldsymbol{k}_\mathrm{o1} + \boldsymbol{k}_\mathrm{o2} + \jj2\omegag\right)s + \jj2\omegag\boldsymbol{k}_\mathrm{o1} = 0.
\end{align}
The poles can be placed at the desired locations through the observer gains. For example, the pole locations can be chosen based on the desired damping ratio $\zetao$ and natural frequency $\omegao$, placing the poles at
\begin{align} \label{eq:pole_placement}
    \boldsymbol{p}_\mathrm{s1,2} = \omegao\left(-\zetao \pm \jj\sqrt{1-\zetao^2}\right) -\jj\omegag.
\end{align}

\subsection{Closed-Loop Control Dynamics} \label{sec:dyn_linear}

\subsubsection{Linearized Model}
Since the grid-voltage estimation errors act like external disturbances to the control system, they do not affect the pole locations and are omitted in the following analysis. The linearized closed-loop control dynamics are derived from the output dynamics \eqref{eq:output_dyn} [which originate from \eqref{eq:model} and \eqref{eq:augmented_model}] and the control law \eqref{eq:gfm_control_law}:  
\begin{subequations} \label{eq:dyn_linear}
\begin{align}
    \frac{\D\Delta\pgp}{\D t} &= \alphap\left(\Delta\pgref - \Delta\pgp\right) + \betapv\left(\Delta\vcref - \Delta\vcpmag\right) \nonumber \\
    &\quad - \RE\left\{\betapn \Delta\vcn \right\} \\
    \frac{\D\Delta\vcpmag}{\D t} &= \alphav\left(\Delta\vcref - \Delta\vcpmag\right) + \betavp\left(\Delta\pgref - \Delta\pgp\right)\nonumber\\
    &\quad - \RE\left\{\betavn\Delta\vcn\right\} \\
    \frac{\D\Delta\vcn}{\D t} &= -\alphan\Delta\vcn + \betanp\left(\Delta\pgref - \Delta\pgp\right) 
    \nonumber\\
    &\quad 
    + \betanv\left(\Delta\vcref - \Delta\vcpmag\right).
\end{align}
\end{subequations}
The coefficients of the positive-sequence active-power and voltage control dynamics are given by  
\begin{subequations} \label{eq:coefficients}
\begin{gather} \nonumber
    \alphap = \frac{1}{L}\RE\left\{\frac{(\ugpo)^*\kpbo}{1 + \jj}\right\} \quad 
    \alphav = \frac{\omegag}{|\vcpo|}\RE\left\{\frac{(\vcpo)^*\kvbo}{1-\jj}\right\} \\ \nonumber
    \betapv = \frac{1}{L}\RE\left\{\frac{(\ugpo)^*\kvbo}{1 + \jj}\right\} \quad 
    \betavp = \frac{\omegag}{|\vcpo|}\RE\left\{\frac{(\vcpo)^*\kpbo}{1-\jj}\right\} \\ 
    \betapn = \frac{1}{L}\frac{\knbo(\ugpo)^*}{1+\jj} \qquad
    \betavn = \frac{\omegag}{|\vcpo|}\frac{\knbo(\vcpo)^*}{1-\jj}. 
\end{gather}
The coefficients of the negative-sequence voltage dynamics are
\begin{gather} \nonumber
    \alphan = \omegag\frac{\knbo + \jj (4 - \knbo)}{2} \\  
    \betanp = \frac{\omegag\kpbo}{1 + \jj} \qquad
    \betanv = \frac{\omegag\kvbo}{1 + \jj} .
\end{gather}
\end{subequations}

The coefficients $\alphap$ and $\alphav$ can be interpreted as the positive-sequence active-power and voltage control bandwidths, respectively. The coefficients $\betapv$ and $\betavp$ represent the coupling between the positive-sequence active-power and voltage control channels, while $\betapn$ and $\betavn$ represent the coupling from the negative-sequence voltage to the positive-sequence active power and voltage, and $\betanp$ and $\betanv$ the reverse coupling. With $\kn > 0$, the negative-sequence voltage feedback enforces $\vcno = 0$. Correspondingly, $\vco = \vcpo$ also holds. 

\subsubsection{Maximum Power-Transfer Boundary}
Based on \eqref{eq:vcp} and \eqref{eq:pg_def}, the positive-sequence active power can be expressed as 
\begin{align} \label{eq:load_angle}
    \pgpo = \frac{1}{\omegag L}\Im{\vcpo(\ugpo)^*} = \frac{|\vcpo| |\ugpo|}{\omegag L}\sin\delta_0 
\end{align}
where $\delta_0 = \angle \vcpo - \angle \ugpo$ is the load angle. Below the maximum power-transfer boundary, the load angle satisfies $-90^\circ < \delta_0 < 90^\circ$, corresponding to the condition $\Re\{\vcpo(\ugpo)^*\} > 0$. This maximum power-transfer condition is independent of the control law and holds for any steady-state operating point.

\subsubsection{Decoupling Gains}
Inserting the gain selection \eqref{eq:gfm_gain_full} into the coefficients in \eqref{eq:coefficients} shows that the positive-sequence control channels are decoupled, i.e., $\betapv=0$ and $\betavp=0$. The corresponding control bandwidths are 
\begin{align} \label{eq:decoupling_full} 
    \alphap = \frac{\kp \Re{\vcpo(\ugpo)^*}}{L\,|\vcpo|} \qquad
    \alphav = \frac{\kv \omegag \Re{\vcpo(\ugpo)^*}}{|\ugpo||\vcpo|}
\end{align}
where $\RE\{\ugpo\vco^*\} > 0$ holds, as mentioned above. 

With the partial decoupling gains \eqref{eq:gfm_gain_partial}, the condition $\betavp = 0$ holds. The active-power-control bandwidth $\alphap$ remains the same as in \eqref{eq:decoupling_full} while the voltage-control bandwidth becomes $\alphav = \kv\omegag/2$. The other coefficients are also obtained from \eqref{eq:coefficients}.

\subsubsection{Stability Conditions}
A case without the negative-sequence feedback, i.e., with $\knbo = 0$, is considered first. With the partial decoupling gains \eqref{eq:gfm_gain_partial}, for which $\betavp=0$, the characteristic polynomial of the fourth-order linearized closed-loop system \eqref{eq:dyn_linear} can be factored as
\begin{equation} \label{eq:char_poly_zero}
    (s + \alphap)(s + \alphav)(s^2 + 4\omegag^2) = 0 .
\end{equation}
The system has stable real poles at $-\alphap$ and $-\alphav$, corresponding to the positive-sequence dynamics. However, the system also has conjugate poles on the imaginary axis at $\pm \jj 2\omegag$. These poles indicate that the system lacks damping for the negative-sequence dynamics, which would result in sustained oscillations at twice the grid frequency in synchronous coordinates.

To introduce the required damping, the negative-sequence feedback gain is chosen as a purely imaginary value $\knbo = \jj\kn$, with $\kn > 0$. This selection introduces the term $-\kn\omegag\hL\icn$ to the control law \eqref{eq:gfm_control_law}, i.e., a virtual resistance for the negative-sequence current. The characteristic polynomial of the general fourth-order system can be written as
\begin{subequations} \label{eq:char_poly} 
\begin{align}
    s^4 + c_3 s^3 + c_2 s^2 + c_1 s + c_0 = 0.
\end{align}
Based on \eqref{eq:dyn_linear}, the coefficients are 
\begin{align}
    c_3 &= \alphap + \alphav + \kn\omegag \\
    c_2 &= \alphap\alphav + \frac{\kn\omegag}{2}\left(\alphap + 2\alphav + \sigma\right) + \left(4 + 2\kn + \frac{\kn^2}{2}\right)\omegag^2 \\
    c_1 &= \omegag^2 \left[4(\alphap + \alphav) + \kn\left(\alphap - \sigma\right) \right] \\ 
    c_0 &= 4\omegag^2\alphap\alphav
\end{align}
\end{subequations} 
where the auxiliary parameter $\sigma = \kp\omegag \pgpo/ |\vcpo|$ is proportional to the positive-sequence active power. 

\begin{figure}[t]
    \centering
    \subfloat[]{\includegraphics[scale=0.45]{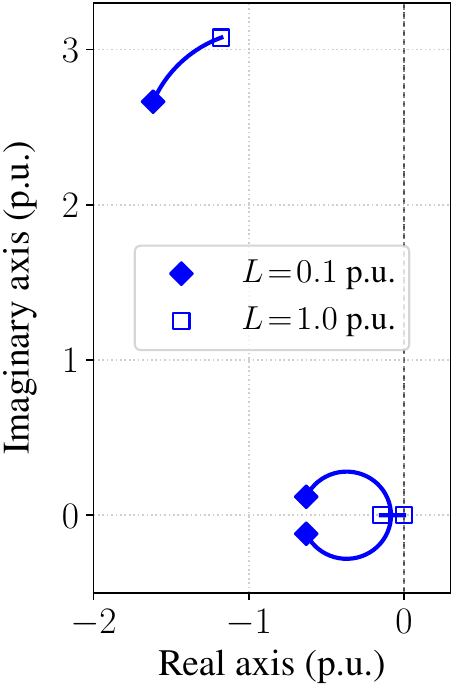}\label{fig:poles1}} \hfil
    \subfloat[]{\includegraphics[scale=0.45]{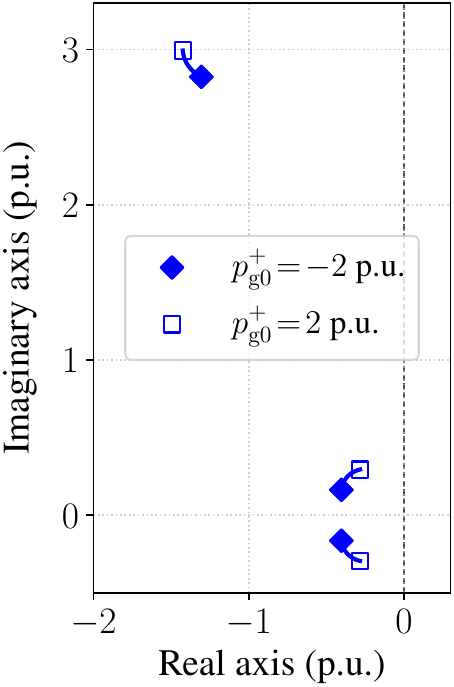}\label{fig:poles2}}
    \caption{Poles of the closed-loop system assuming an accurate inductance estimate: (a) inductance $L$ varies between 0.1 and 1 p.u.; (b) active power $\pgpo$ varies between $-$2 and 2 p.u. The converter-voltage magnitude $\vcmago$, grid-voltage magnitude $\ugmago$, and grid angular frequency $\omegag$ are all $1$ p.u. In (a), the active power is $\pgpo=1$ p.u., while in (b) the inductance is $L=0.2$ p.u. The lower half-plane is only partly shown due to pole symmetry.}
    \label{fig:poles}
\end{figure}

The stability conditions are obtained by applying the Routh--Hurwitz criterion to the characteristic polynomial \eqref{eq:char_poly}. First, all coefficients must be positive. An additional stability condition $c_3 c_2 c_1 > c_1^2 + c_3^2 c_0$ originates from the Routh array. A necessary stability condition originates from $c_1 > 0$. In the inverter mode at high active power ($\sigma > \alphap$), the imaginary gain is upper-bounded by
\begin{equation} \label{eq:kn_limit}
    \kn < \frac{4(\alphap + \alphav)}{\sigma - \alphap} .
\end{equation}
This stability condition means that the gain $\kn$ has a strict upper limit only when operating in the inverter mode at high active power. In the rectifier mode, or at a lower active power in the inverter mode ($\sigma < \alphap$), the gain $\kn$ is not limited by this condition. Thus, the proposed imaginary gain allows a wide stable operating region. With the parameters of Table~\ref{tab:param_ctrl} and $|\vcpo| = |\ugpo| = 1$~p.u., all stability conditions are fulfilled at every operating point below the maximum power-transfer boundary where the converter current does not exceed $i_\mathrm{max}$. Hence, \eqref{eq:kn_limit} does not limit the operating region in practice.

To illustrate the stability of the closed-loop system, the poles are numerically calculated. Fig.~\ref{fig:poles} shows the poles as the inductance and operating-point active power vary. The grid-voltage magnitude and frequency are $\ugmago=1$~p.u. and $\omegag=1$~p.u., respectively, and the converter-voltage magnitude is $\vcmago=1$~p.u. The control parameters are $\kp=0.2$~p.u., $\kv=1$, and $\kn=2$, and the partial decoupling gains \eqref{eq:gfm_gain_partial} are used. The selection of $\kp$ and $\kv$ is based on the tuning guidelines presented in~\cite{Nur2024}. The same tuning is used in the experiments. 

Fig.~\ref{fig:poles}(a) shows the poles while the inductance $L$ varies between 0.1 and 1 p.u. The active power is $\pgpo=1$ p.u. As the inductance increases, the system approaches the stability limit and becomes marginally stable exactly at $L=1$ p.u. This coincides with the theoretical limit for the maximum power transfer. Fig.~\ref{fig:poles}(b) shows the poles while the active power varies between $-2$ and $2$ p.u., covering both the rectifier and inverter modes. For this analysis, current limits are ignored. The inductance is $L=0.2$ p.u. The operating point clearly affects the pole locations, but they remain in the left half-plane. These results confirm that the system maintains stability across a wide range of operating conditions.

\subsection{Effect of Inductance Estimation Error} \label{sec:inductance_error}

\subsubsection{Estimation Errors in Steady State}

Using \eqref{eq:model}, \eqref{eq:obs_cont}, \eqref{eq:gfm_control_law}, and the estimation-error dynamics from Appendix~\ref{app:inductance_error}, the grid-voltage estimation errors in the periodic steady state are
\begin{subequations}
\begin{align}
    \tildeugpo &= - \jj\omegag\tL\icpo\label{eq:tildeugpo}\\
    \tildeugno &= \jj\omegag\tL\icno.
\end{align}
\end{subequations}

\begin{subequations} \label{eq:ctrl_est}
Considering the estimation error for the positive-sequence grid voltage due to the inductance estimation error, the estimates for the positive-sequence active power and voltage magnitude are
\begin{align}
    \hatpgpo &= \pgpo - \Re{\tildeugpo(\icpo)^*}\\
    \hvcpmago &= |\vcpo - \tildeugpo - \jj\omegag\tL\icpo|.
\end{align}
\end{subequations}
Substituting \eqref{eq:tildeugpo} into \eqref{eq:ctrl_est} results in
\begin{subequations}
\begin{align}
    \hatpgpo &= \pgpo\\
    \hvcpmago &= |\vcpo|.
\end{align}
\end{subequations}
The inductance estimation error introduces an operating-point-dependent error in the estimated grid voltage. However, the estimates of the controlled variables stay exact, i.e., the inductance estimation error does not cause steady-state errors in active power and quasi-static converter voltage.

\subsubsection{Numerical Analysis}

\begin{figure}[t]
    \centering
    \subfloat[]{\includegraphics[scale=0.45]{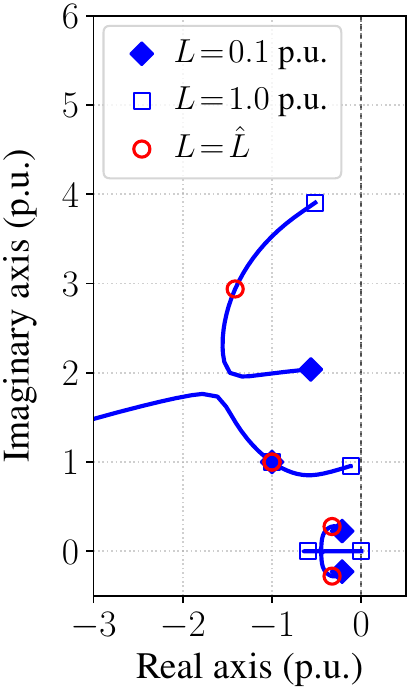}} \hfil
    \subfloat[]{\includegraphics[scale=0.45]{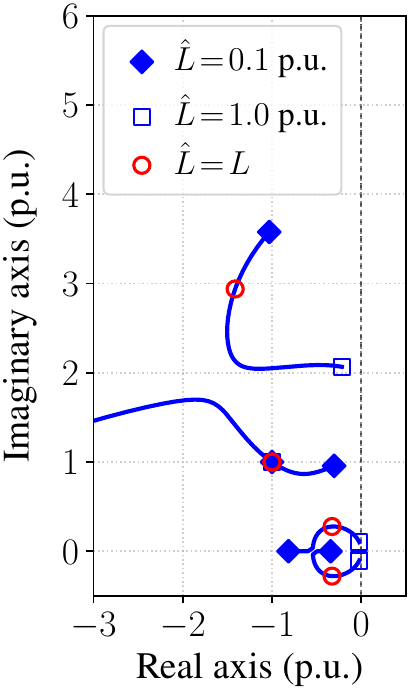}}
    \caption{Poles of the coupled closed-loop system considering the inductance estimation error: (a) actual inductance $L$ varies between 0.1 and 1~p.u. while the inductance estimate is $\hL=0.2$~p.u.; (b) inductance estimate $\hL$ varies between 0.1 and 1~p.u. while the actual inductance is $L=0.2$~p.u. The active power is $\pgpo=1$~p.u. The red circles mark $\hL=L$. The lower half-plane is only partly shown due to pole symmetry. In (a), the fastest pole pair leaves the plotted area and reaches $-4.12\pm\jj1.14$ for $L=0.1$~p.u. In (b), the fastest pole pair splits into two real poles, located at $-10.2$ and $-8.0$ for $\hL=1$~p.u.}
    \label{fig:poles_coupled_L}
\end{figure}

The poles of the coupled eighth-order system are computed numerically. Fig.~\ref{fig:poles_coupled_L} shows the poles as the actual inductance and the inductance estimate vary. The full model is derived in Appendix~\ref{app:inductance_error}. The observer poles are placed using \eqref{eq:pole_placement} with $\zetao=1$ and $\omegao=1$~p.u., while other parameters are the same as in Fig.~\ref{fig:poles}.

Fig.~\ref{fig:poles_coupled_L}(a) shows the poles as the actual inductance varies between 0.1 and 1~p.u. while the inductance estimate is kept at $\hL=0.2$~p.u. The active power is $\pgpo=1$~p.u. All poles remain in the left half-plane for $L<1$~p.u., showing that the closed-loop system remains stable up to the maximum power-transfer boundary even with severe underestimation of the inductance. Fig.~\ref{fig:poles_coupled_L}(b) shows the poles as the inductance estimate varies between 0.1 and 1~p.u. while the actual inductance is $L=0.2$~p.u. The active power is $\pgpo=1$~p.u. All poles remain in the left half-plane for $\hL<1$~p.u., showing that the proposed controller also tolerates severe overestimation of the inductance.

The same model shows that the operating point affects the poles less than the inductance estimation error: with $\hL = 0.2$~p.u., the system remains stable for $\pgpo$ from $-2$ to $2$~p.u. at $L = 0.1$~p.u. and within the maximum power-transfer boundary at $L = 1$~p.u. This demonstrates that stability is maintained across a wide range of operating conditions even with severe inductance estimation errors.

\section{Implementation Aspects} \label{chap:implementation}

\subsection{Reference Limitation}
Reference limitation is used together with transparent current control to keep the converter current within allowable limits. The aim is to maintain grid-forming operation, i.e., positive-sequence voltage-source behavior, and to permit the flow of negative-sequence current during unbalanced faults. This is achieved by limiting the active-power reference to a realizable level during faults and by employing an elliptical current limiter.

\subsubsection{Active-Power Reference}
To prioritize reactive power and negative-sequence current during faults, the maximum allowable active power $p_\mathrm{max}$ is first computed as
\begin{subequations}\label{eq:powerlim}
\begin{align}
    i_\mathrm{q}^+ &= \IM\left\{\frac{\ucrefp(\icp)^*}{|\ucrefp|}\right\} \\
    i_\mathrm{d,lim}^2 &= \max\left(i_\mathrm{d,max} - |\icn|,\ 0\right)^2 - \bigl(i_\mathrm{q}^+\bigr)^2 \\
    p_\mathrm{max} &= \frac{\alpha_\mathrm{l}}{s + \alpha_\mathrm{l}}|\ucrefp|\sqrt{\max\bigl(i_\mathrm{d,lim}^2,\ 0\bigr)}
\end{align}
where $i_\mathrm{q}^+$ is the positive-sequence reactive current, $i_\mathrm{d,max}$ is the maximum active current, and $\alpha_\mathrm{l}$ is the power-limitation bandwidth. The limited active-power reference is given by
\begin{align}
    \pgreflim &= \SIGN(\pgref)\min(|\pgref|,\ p_\mathrm{max}).
\end{align}
\end{subequations}

\subsubsection{Current Reference}
An elliptical current limiter \cite{Zarei2019,Awal2023} is used to limit the steady-state converter phase currents to the maximum allowable value. First, the maximum phase current $i_\mathrm{ph,max}$ is computed from the unlimited current references as 
\begin{subequations}\label{eq:curlim}
\begin{align}
    i_\mathrm{ph,max} = \max_{n \in \{-1, 0, 1\}}\left|\icref^\pl + (\icref^\mi)^*\mathrm{e}^{\mathrm{j}(2n\pi/3 - 2\thetac)}\right|.
\end{align}
The positive- and negative-sequence current references are then separately limited as
\begin{align}
    \begin{bmatrix}
        \icreflim^\pl\\ \icreflim^\mi
    \end{bmatrix}
    = \frac{i_\mathrm{max}}{\max\left(i_\mathrm{ph,max},i_\mathrm{max}\right)}
    \begin{bmatrix}
        \icref^\pl\\ \icref^\mi
    \end{bmatrix}
\end{align}
\end{subequations}
where $i_\mathrm{max}$ is the maximum allowable steady-state converter phase current. The elliptical trajectory of the current-reference vector is preserved by scaling both sequence components with the same dynamic factor. Consequently, no additional current harmonics are injected into the grid.

\begin{figure}[t] 
    \centering
    \subfloat[]{\includegraphics{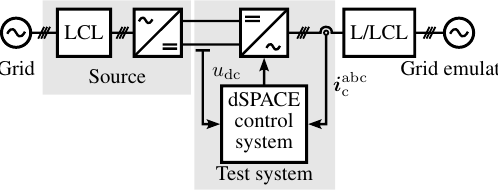}} \\
    \subfloat[]{\includegraphics[align=c,width=0.75\columnwidth]{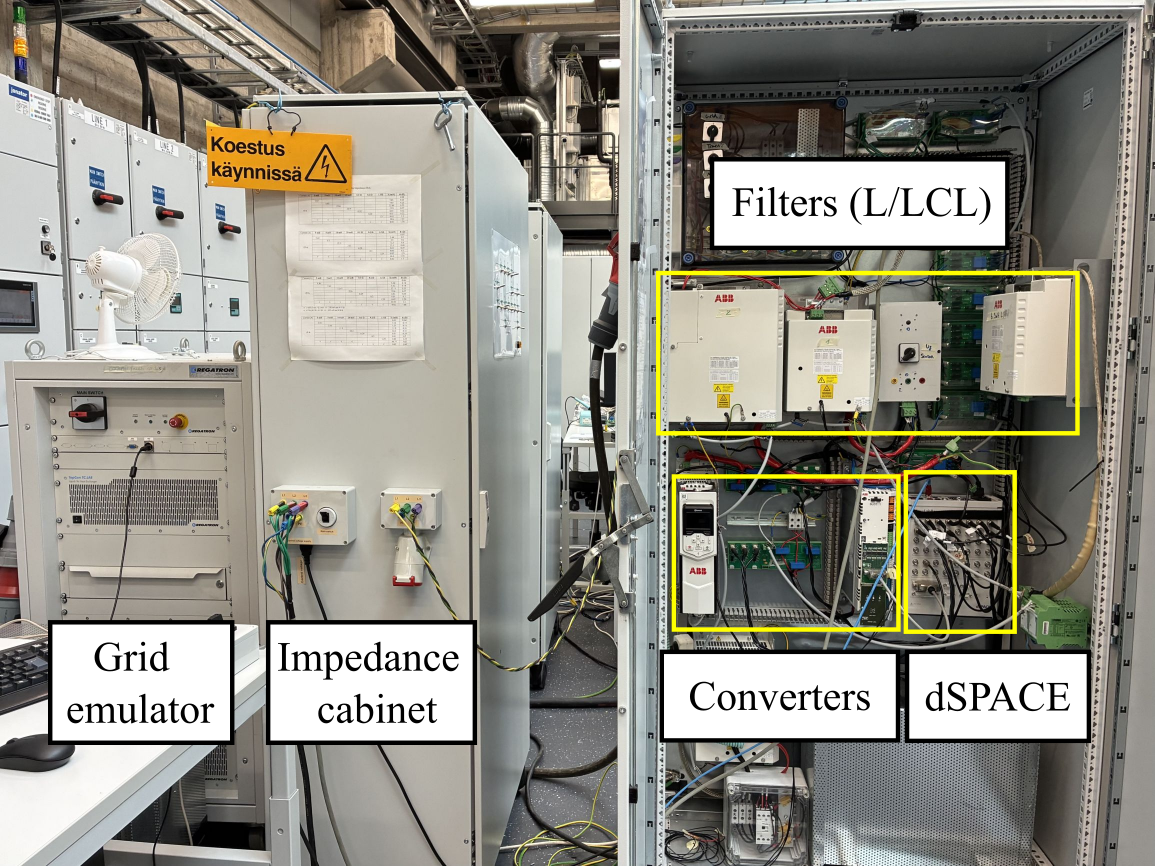}}
    \caption{Experimental setup: (a) circuit diagram; (b) photograph. The impedance cabinet contains additional inductors for emulating weak grids.}
    \label{fig:setup}
\end{figure}

\subsection{Experimental Setup}
Fig.~\ref{fig:setup} shows the laboratory setup and Table~\ref{tab:param_grid} provides its parameters. The experimental setup comprises two back-to-back 12.5-kVA three-phase converters and a 50-kVA four-quadrant programmable grid emulator (Regatron TopCon TC.ACS). The test converter can be equipped with either an L or an LCL filter. An impedance cabinet contains additional inductors for emulating various grid strengths. A dSPACE MicroLabBox prototyping unit controls the converter and gathers measurement data. The grid voltage is measured only for monitoring purposes and for computing $\pg$ and $\qg$ shown in the results of Section~\ref{chap:results}.

The proposed control method comprises the control system \eqref{eq:obs_cont}--\eqref{eq:CC_mode}, the active-power limitation \eqref{eq:powerlim}, and the current limitation \eqref{eq:curlim}. The disturbance observer is implemented according to \eqref{eq:disc_obs}. The corresponding continuous-time observer poles are placed using \eqref{eq:pole_placement}.
In place of the delayed voltage reference \eqref{eq:delay}, the realized converter voltage $\bar{\boldsymbol{u}}_\mathrm{c,ref}$, computed from the duty ratios and measured DC-bus voltage, is used in the observer.
The one-sampling-period computational time delay and half-sampling-period pulse-width modulation time delay are compensated for by transforming the converter-voltage reference to stationary coordinates as
\begin{align}
    \ucsref = \e^{\jj\thetac}\left(\e^{\jj3\omegag\Ts/2}\ucrefp + \e^{-\jj3\omegag\Ts/2}\ucrefn\right).
\end{align}

Table~\ref{tab:param_ctrl} lists the control gains and other controller parameters. Experiments are conducted with both an L and an LCL filter, and the same inductance estimate is used for both filters. The same inductance estimate is also used in the strong and weak grids to study the effects of inductance estimation errors. In the experiments using the L filter, the value of $L$ refers to the total inductance comprising the filter and grid inductances. With the LCL filter, the value of $\Lg$ refers to the sum of the grid-side filter inductance and grid inductance.

Switching frequencies of $4$~kHz and $8$~kHz are used with the L and LCL filters, respectively. With the LCL filter, the higher switching frequency is chosen to increase the margin between the switching frequency and the resonance frequency of the LCL filter, which is around 1.3~kHz. In both cases, the sampling frequency is twice the switching frequency.

\begin{table}[t] \small \centering
\caption{Experimental Setup Data} \label{tab:param_grid}
\begin{tabular}{lcc}
\toprule
Parameter & Actual value & Per-unit value\\
\midrule
Rated voltage & $\sqrt{2/3}\cdot400$ V\! & 1 p.u.\\
Rated current & $\sqrt{2}\cdot18$ A & 1 p.u. \\
Rated power & $12.5$ kVA & 1 p.u. \\
Rated frequency & $50$ Hz & 1 p.u.\\
DC-bus voltage &$660$ V & 2 p.u.\\
Switching frequency & $4/8$ kHz & $80/160$ p.u. \\
Sampling frequency & $8/16$ kHz & $160/320$ p.u. \\
L filter inductance & $6.3$ mH & 0.15 p.u.\\
LCL converter-side inductance  & $3.3$ mH & 0.08 p.u.\\
LCL grid-side inductance & $3.0$ mH & 0.07 p.u.\\
LCL capacitance & $8.8$ µF & 0.04 p.u.\\
\bottomrule 
\end{tabular}
\end{table}

\begin{table}[t] \small \centering
\caption{Controller Parameters} \label{tab:param_ctrl}
\begin{tabular}{lcc}
\toprule
Parameter & Symbol & Value\\
\midrule
Maximum current & $i_\mathrm{max}$ & 1.3 p.u.\\
Maximum active current & $i_\mathrm{d,max}$ & 1.1 p.u.\\
Power-control gain & $\kp$ & 0.2 p.u.\\
Voltage-control gain & $\kv$ & 1 \\
Negative-sequence voltage gain & $\kn$ & 2 \\
Inductance estimate & $\hat L$ & 0.2 p.u.\\
Current-control bandwidth & $\alpha_\mathrm{c}$ & 6 p.u.\\
Power-limitation bandwidth & $\alpha_\mathrm{l}$ & 1 p.u.\\
Observer damping ratio & $\zeta_\mathrm{o}$ & 1\\
Observer natural frequency & $\omega_\mathrm{o}$ & 1 p.u.\\
\bottomrule 
\end{tabular}
\end{table}

\section{Experimental Results} \label{chap:results}
Experimental tests are conducted to verify the operation of the proposed control method under unbalanced and balanced faults and during active-power reference tracking. Simulation models for the same test cases are available in \cite{motulator_capsule}.

\subsection{Baseline Without Sequence Separation}
\begin{figure}[t]
    \centering
    \includegraphics[width=0.95\columnwidth]{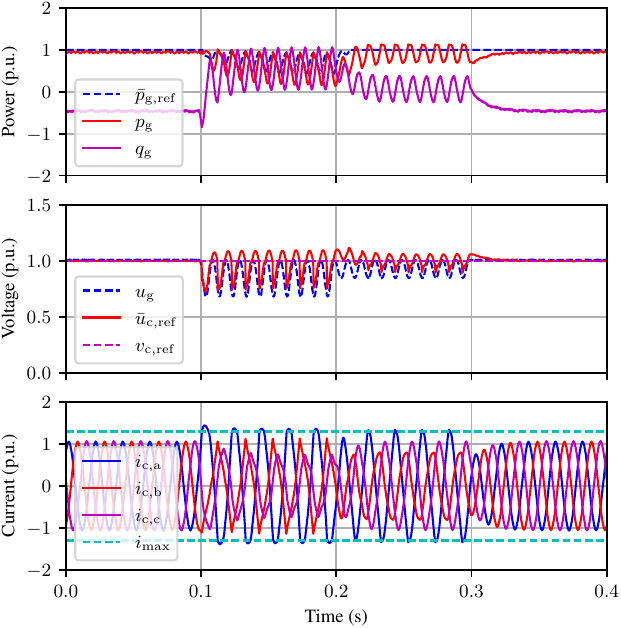}
    \caption{Experimental results for the method in \cite{Nur2024} with an L filter for an unbalanced fault in a strong grid ($L=0.15$ p.u.). Two different fault levels with the phase-a grid-voltage magnitude at $u_\mathrm{g,a}=0.5$ p.u. and $u_\mathrm{g,a}=0.75$ p.u. are shown.}
    \label{fig:do_gfm_comparison}
\end{figure}

Fig.~\ref{fig:do_gfm_comparison} shows the grid-forming control method in~\cite{Nur2024} during an unbalanced fault. The three-phase quantities are marked with the subscripts a, b, and c. A single-phase fault is generated by lowering the phase-a grid-voltage magnitude to $u_\mathrm{g,a}=0.5$~p.u. at $t=0.1~\mathrm{s}$. The voltage magnitude is raised to $u_\mathrm{g,a}=0.75$~p.u. at $t=0.2~\mathrm{s}$ and to $u_\mathrm{g,a}=1$~p.u. at $t=0.3~\mathrm{s}$. An L filter and a strong grid ($L=0.15$ p.u.) are considered. The method implements only positive-sequence control and, thus, during the unbalanced fault the converter output currents are nonsinusoidal and the current limit is violated. Furthermore, when the phase-a voltage rises to 0.75~p.u., the converter output voltage remains unbalanced and the current limit is reached for phase a.

To further study the nonsinusoidal output currents, the total harmonic distortion (THD) is computed. In a separate test, the phase-a grid-voltage magnitude is lowered to $u_\mathrm{g,a}=0.5$~p.u. for 1 s. The fast Fourier transform of a ten-period window is used to obtain the harmonic components up to the Nyquist frequency of 4 kHz. The THD values for the converter currents are 8.6\%, 15\%, and 13\% for phases a, b, and c, respectively. These THD values significantly exceed 5\%, which is used as a reference value.

\subsection{Unbalanced Fault}
\begin{figure}[t]
    \centering
    \subfloat[]{\includegraphics[width=0.95\columnwidth]{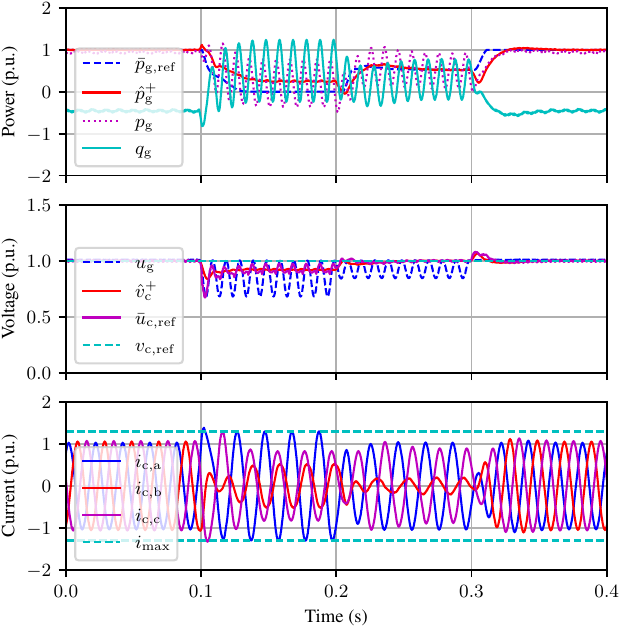}} \\
    \subfloat[]{\includegraphics[width=0.95\columnwidth]{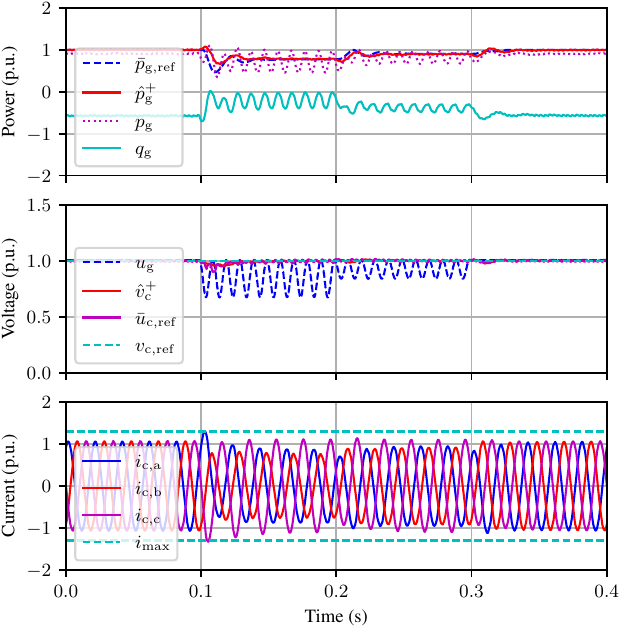}}
    \caption{Experimental results for the proposed method with an L filter for an unbalanced fault in (a) strong grid ($L=0.15$ p.u.) and (b) weak grid ($L=0.77$ p.u.). Two different fault levels with the phase-a grid-voltage magnitude at $u_\mathrm{g,a}=0.5$ p.u. and $u_\mathrm{g,a}=0.75$ p.u. are shown.}
    \label{fig:fault_asym}
\end{figure}

\begin{figure}[t]
    \centering
    \subfloat[]{\includegraphics[width=0.95\columnwidth]{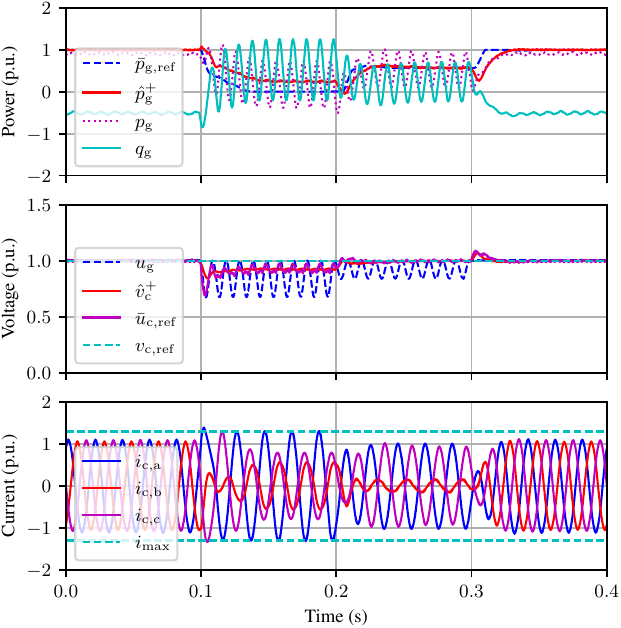}} \\
    \subfloat[]{\includegraphics[width=0.95\columnwidth]{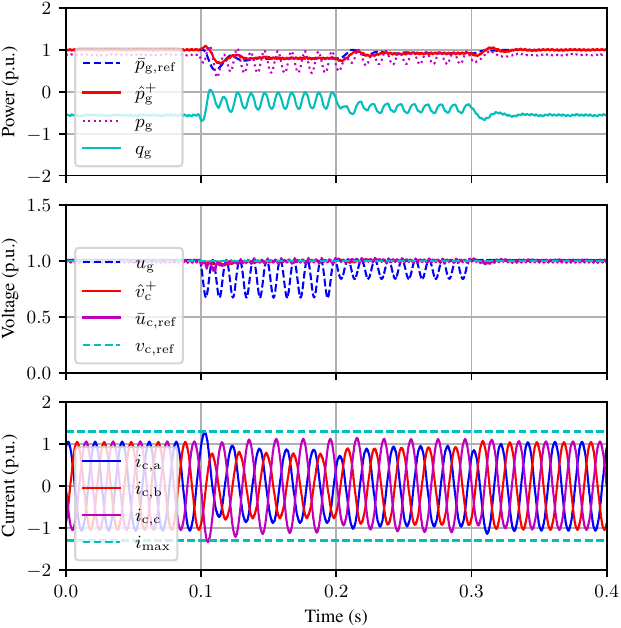}}
    \caption{Experimental results for the proposed method with an LCL filter for an unbalanced fault in (a) strong grid ($\Lg=0.07$ p.u.) and (b) weak grid ($\Lg=0.7$ p.u.). Two different fault levels with the phase-a grid-voltage magnitude at $u_\mathrm{g,a}=0.5$ p.u. and $u_\mathrm{g,a}=0.75$ p.u. are shown.}
    \label{fig:fault_asym2}
\end{figure}

Figs.~\ref{fig:fault_asym} and~\ref{fig:fault_asym2} show the proposed control method during the same unbalanced fault as in Fig.~\ref{fig:do_gfm_comparison} with the L and the LCL filter, respectively. During the deeper fault in a strong grid, the current-control mode is activated and the maximum phase-current magnitude is limited to the maximum value in the periodic steady state, while sinusoidal currents are maintained. A purely positive-sequence converter voltage cannot be maintained because of the phase-current limit. This can be seen as an oscillation in the converter-voltage magnitude $\ucreflim$. When the phase-a voltage rises to 0.75~p.u., the converter-voltage magnitude remains constant. In Fig.~\ref{fig:do_gfm_comparison}, by contrast, the converter-voltage magnitude oscillates throughout the fault.

The THD values corresponding to Fig.~\ref{fig:fault_asym}(a), computed from a ten-period current window during a longer 0.5~p.u. fault in a strong grid, are 1.6\%, 4.9\%, and 0.88\% for phases a, b, and c, respectively. These values are all within the 5\% reference value, showing that during unbalanced conditions the proposed method achieves significantly better power quality than the baseline method.

In a weak grid, grid-forming operation continues after the initial transient at $t=0.1~\mathrm{s}$ and a constant converter-voltage magnitude is maintained in the periodic steady state. This demonstrates that a purely positive-sequence converter voltage can be maintained during unbalanced faults as well, provided that the current limit is not reached. The results obtained with the L and LCL filters appear to be nearly identical.

\subsection{Balanced Fault}
\begin{figure}[t]
    \centering
    \subfloat[]{\includegraphics[width=0.95\columnwidth]{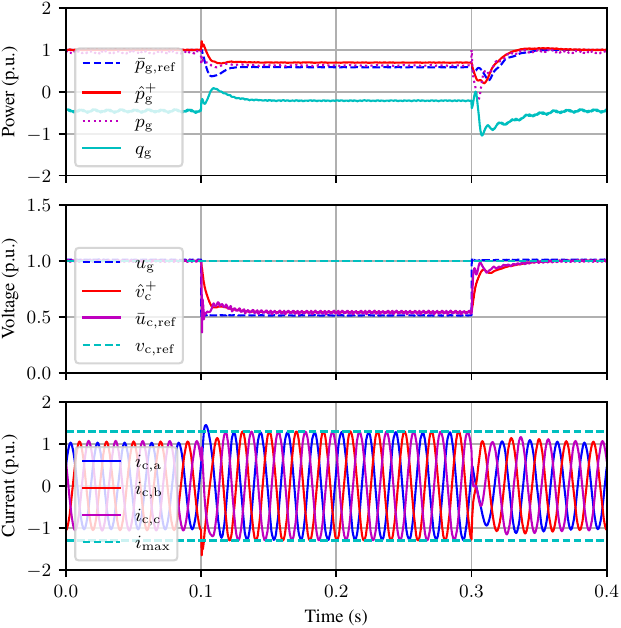}} \\  
    \subfloat[]{\includegraphics[width=0.95\columnwidth]{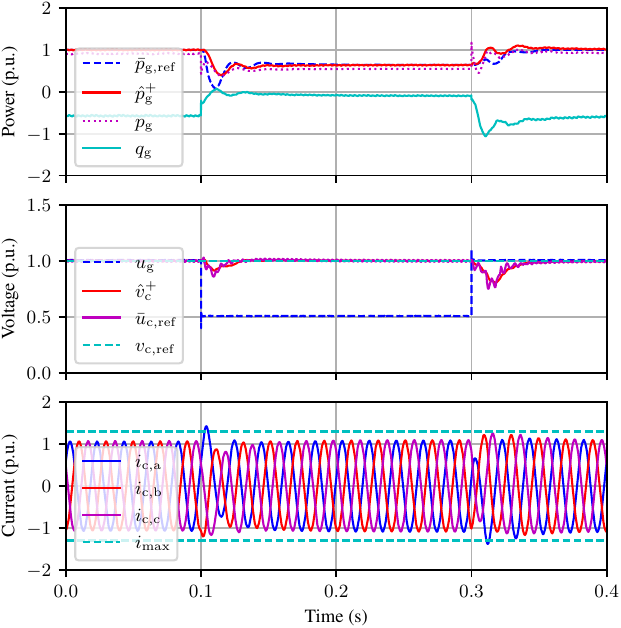}}
    \caption{Experimental results for the proposed method with an L filter for a balanced fault in (a) strong grid ($L=0.15$ p.u.) and (b) weak grid ($L=0.77$ p.u.). The grid-voltage magnitude is $\ugmag=0.5$ p.u. during the fault.}
    \label{fig:fault_sym}
\end{figure}

\begin{figure}[t]
    \centering
    \subfloat[]{\includegraphics[width=0.95\columnwidth]{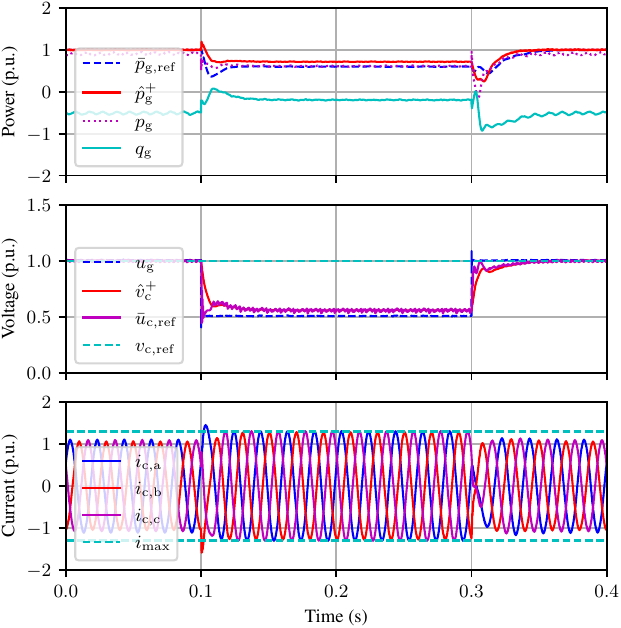}} \\
    \subfloat[]{\includegraphics[width=0.95\columnwidth]{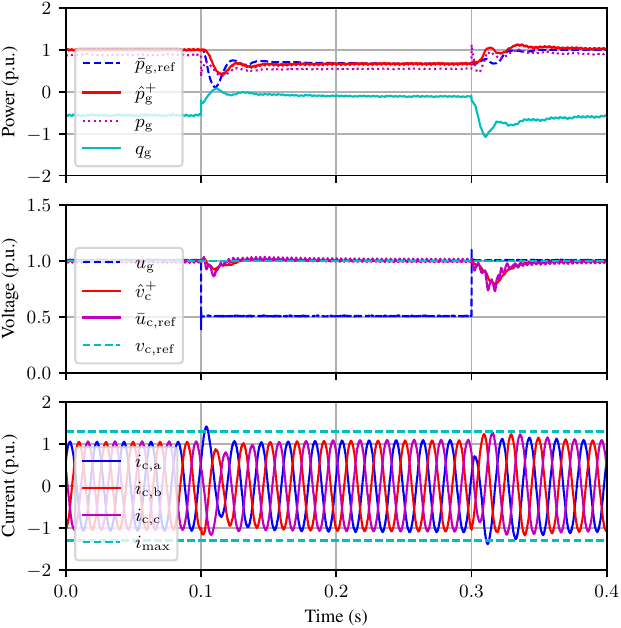}}
    \caption{Experimental results for the proposed method with an LCL filter for a balanced fault in (a) strong grid ($\Lg=0.07$ p.u.) and (b) weak grid ($\Lg=0.7$ p.u.). The grid-voltage magnitude is $\ugmag=0.5$ p.u. during the fault.}
    \label{fig:fault_sym2}
\end{figure}

Figs.~\ref{fig:fault_sym} and~\ref{fig:fault_sym2} show the proposed control method during a balanced fault with the L and the LCL filter, respectively. The grid-voltage magnitude is lowered from 1~p.u. to 0.5~p.u. at $t=0.1~\mathrm{s}$ and raised back to 1~p.u. at $t=0.3~\mathrm{s}$. In a strong grid, a constant converter voltage cannot be maintained since this would imply very large currents. Thus, the converter remains in the current-control mode for the duration of the fault, after which grid-forming operation is resumed. The offset between the active-power reference and the positive-sequence active power occurs because the converter is in the current-control mode.

In a weak grid, the converter switches back to the grid-forming mode when the reference limiter provides a realizable active-power reference, thus supporting the system voltage while maintaining the current magnitude within permissible limits. Once again, there is no discernible difference between the results obtained with the L and LCL filters.

\subsection{Active-Power Reference Tracking}
\begin{figure}[t]
    \centering
    \subfloat[]{\includegraphics[width=0.95\columnwidth]{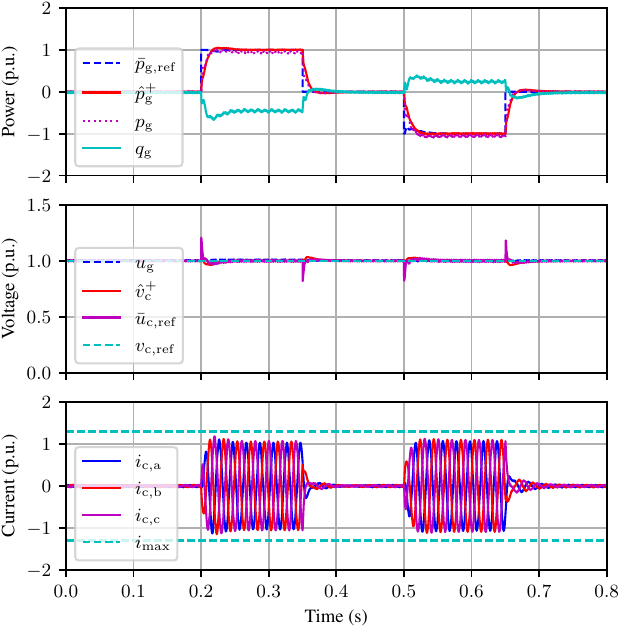}} \\
    \subfloat[]{\includegraphics[width=0.95\columnwidth]{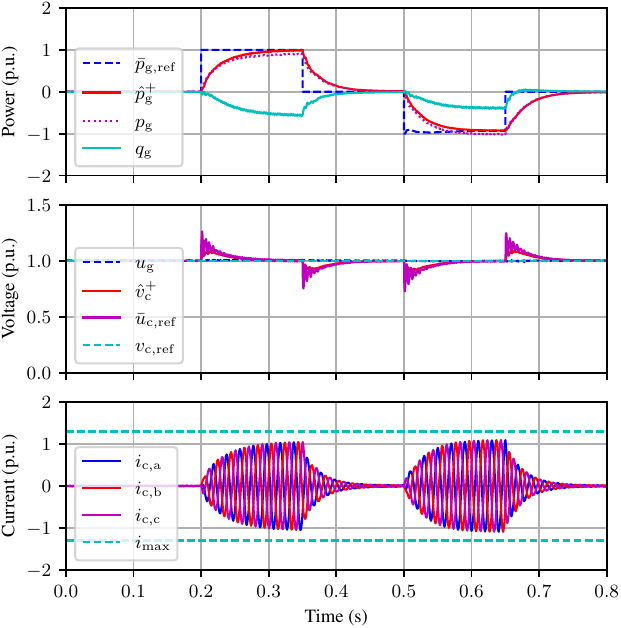}}
    \caption{Experimental results for the proposed method with an L filter and active-power reference step changes in (a) strong grid ($L=0.15$ p.u.) and (b) weak grid ($L=0.77$ p.u.). The active-power reference is changed stepwise: $\pgref=1$ p.u. at $t=0.2~\mathrm{s}$; $\pgref=0$ p.u. at $t=0.35~\mathrm{s}$; $\pgref=-1$ p.u. at $t=0.5~\mathrm{s}$; $\pgref=0$ p.u. at $t=0.65~\mathrm{s}$. The limited active-power reference $\pgreflim$ is shown.}
    \label{fig:steps}
\end{figure}

\begin{figure}[t]
    \centering
    \subfloat[]{\includegraphics[width=0.95\columnwidth]{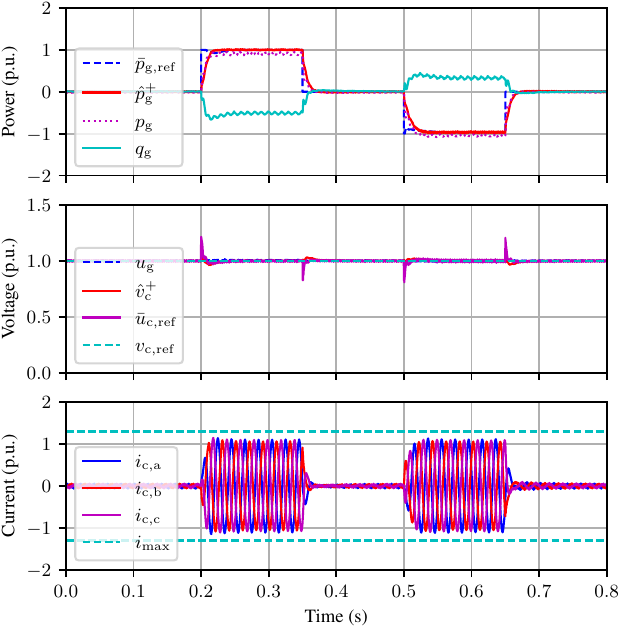}} \\
    \subfloat[]{\includegraphics[width=0.95\columnwidth]{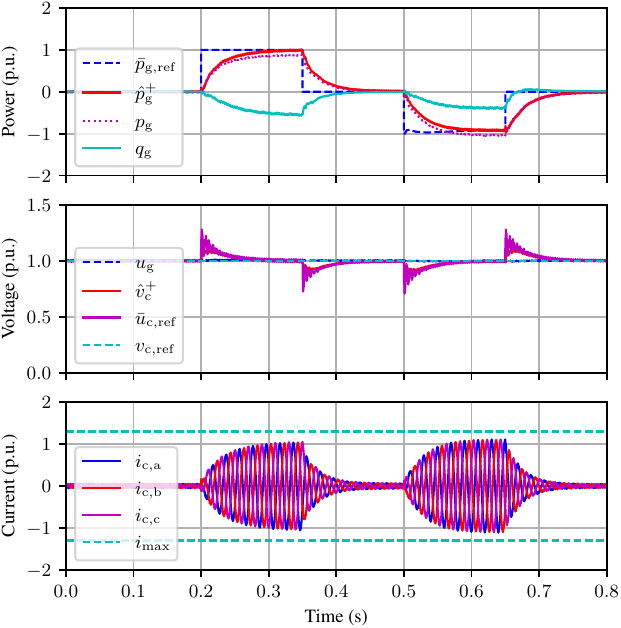}}
    \caption{Experimental results for the proposed method with an LCL filter and active-power reference step changes in (a) strong grid ($\Lg=0.07$ p.u.) and (b) weak grid ($\Lg=0.7$ p.u.). The active-power reference is changed stepwise: $\pgref=1$ p.u. at $t=0.2~\mathrm{s}$; $\pgref=0$ p.u. at $t=0.35~\mathrm{s}$; $\pgref=-1$ p.u. at $t=0.5~\mathrm{s}$; $\pgref=0$ p.u. at $t=0.65~\mathrm{s}$. The limited active-power reference $\pgreflim$ is shown.}
    \label{fig:steps2}
\end{figure}

Figs.~\ref{fig:steps} and~\ref{fig:steps2} show active-power reference tracking for the proposed method with the L and the LCL filter, respectively. The active-power reference is changed in steps between 1 and $-$1~p.u. The limited active-power reference $\pgreflim$ is also shown. There is a slight overshoot in the active power in a strong grid with the L filter but not with the LCL filter. With the LCL filter, the offset between the positive-sequence active-power estimate $\hatpgp$ and the active power $\pg$ computed from the measured grid voltage is slightly larger than in the L-filter case. Both effects can be attributed to the LCL filter having a slightly higher resistance than the L filter. The proposed method achieves good active-power reference tracking in both strong and weak grids while the inductance estimate remains the same in both cases, showing excellent robustness against inductance estimation error.

\section{Conclusion} \label{chap:conclusion}
A grid-forming control method for unbalanced grid-voltage conditions was proposed. The method uses a disturbance observer to unify synchronization and estimation of the positive and negative sequences of the grid voltage, requiring no AC-side voltage measurement. Additionally, a current-control mode was introduced to ensure safe operation during faults. A linearized small-signal model of the control system was derived to show the stability of the proposed method over a wide operating region.

Experimental results demonstrate that the proposed method successfully maintains the desired positive-sequence converter voltage whenever allowed by the physical limits of the converter and respects the current limits under both balanced and unbalanced faults. Furthermore, the active-power reference is tracked reliably in strong and weak grids, exhibiting excellent robustness against severe inductance estimation errors. Finally, the control method is shown to function effectively with an LCL filter, despite the control design being based on a simpler L-filter model.

\appendices
\section{Direct Discrete-Time Design} \label{app:zoh}
First, the exact zero-order-hold equivalent model is derived. Then, the discrete-time variant of the disturbance observer \eqref{eq:obs_cont} is designed, using the approach in \cite{Fra1997}. An exact discrete-time variant of the sequence-separation filter \eqref{eq:sequence_filter} is also given.

\subsection{Exact Hold-Equivalent Model}
The hold-equivalent model of \eqref{eq:model} is 
\begin{subequations} \label{eq:disc}
\begin{align}
    \ic(k\!+\!1) &= \boldsymbol{\phi}\ic(k) + \boldsymbol{\gamma}_\mathrm{c}\uc(k)\nonumber\\ &\quad + \boldsymbol{\gamma}_\mathrm{g}^\pl\ugp(k) + \boldsymbol{\gamma}_\mathrm{g}^\mi\ugn(k) \\
    \!\ugp(k\!+\!1) &= \boldsymbol{\phi}_\mathrm{g}^\pl \ugp(k) \\
    \!\ugn(k\!+\!1) &= \boldsymbol{\phi}_\mathrm{g}^\mi \ugn(k)
\end{align}
where $\boldsymbol{\phi} = \mathrm{e}^{-\jj\omegag\Ts}$, $\boldsymbol{\phi}_\mathrm{g}^\pl = 1$, and $\boldsymbol{\phi}_\mathrm{g}^\mi = \boldsymbol{\phi}^2$ are the state-transition coefficients, and $\Ts$ is the sampling period. A constant $\omegac=\omegag$ is assumed. The input coefficient for the converter voltage is
\begin{align} 
    \boldsymbol{\gamma}_\mathrm{c} = \frac{\Ts \boldsymbol{\phi}}{L}
\end{align}
where the converter voltage $\uc$ is assumed constant over the sampling period in stationary coordinates. The input coefficient for the positive-sequence voltage is
\begin{align} 
    \boldsymbol{\gamma}_\mathrm{g}^\pl = \frac{\boldsymbol{\phi} - 1}{\jj\omegag L} 
\end{align}
where the voltage $\ugp$ is assumed to rotate at a constant angular speed $\omegag$ with respect to stationary coordinates over the sampling period. The input coefficient for the negative-sequence voltage is
\begin{align} 
    \boldsymbol{\gamma}_\mathrm{g}^\mi = \boldsymbol{\phi}\boldsymbol{\gamma}_\mathrm{g}^\pl
\end{align}
\end{subequations}
where the voltage $\ugn$ is assumed to rotate at a constant angular speed $-\omegag$ with respect to stationary coordinates over the sampling period. Finally, the model for the one-sampling-period computational delay is 
\begin{align} \label{eq:delay}
    \uc(k) = \boldsymbol{\phi}\ucref(k-1).
\end{align}

\subsection{Disturbance Observer}
Based on the hold-equivalent model \eqref{eq:disc}, the discrete-time variant of the disturbance observer \eqref{eq:obs_cont} is formulated as 
\begin{subequations} \label{eq:disc_obs}
\begin{align}
    \hatugp(k) &= \boldsymbol{\phi}_\mathrm{g}^\pl \hatugp(k-1) + \ko^\pl\eo(k)\\
    \hatugn(k) &= \boldsymbol{\phi}_\mathrm{g}^\mi \hatugn(k-1) + \ko^\mi\eo(k)
\end{align}
where $\ko^\pl$ and $\ko^\mi$ are the discrete-time observer gains. The correction term is
\begin{align} \label{eq:disc_eo}
    \eo(k) &= \ic(k) - \boldsymbol{\phi}\ic(k-1) - \hat{\boldsymbol{\gamma}}_\mathrm{c}\uc(k-1)\nonumber\\
    &\quad - \hat{\boldsymbol{\gamma}}_\mathrm{g}^\pl\hatugp(k-1) - \hat{\boldsymbol{\gamma}}_\mathrm{g}^\mi\hatugn(k-1)
\end{align}
where the hats on the input coefficients indicate that they are computed based on the inductance estimate $\hat L$. The computational delay is taken into account 
using \eqref{eq:delay}.
    
From \eqref{eq:disc} and \eqref{eq:disc_obs}, the second-order characteristic equation for the estimation dynamics can be derived. The observer gains $\ko^\pl$ and $\ko^\mi$ can be computed as
\begin{align}
    \ko^\pm = \frac{(\boldsymbol{\phi}_\mathrm{g}^\pm - \boldsymbol{p}_\mathrm{z1})(\boldsymbol{\phi}_\mathrm{g}^\pm - \boldsymbol{p}_\mathrm{z2})}{\hat{\boldsymbol{\gamma}}_\mathrm{g}^\pm(\boldsymbol{\phi}_\mathrm{g}^\pm - \boldsymbol{\phi}_\mathrm{g}^\mp)}
\end{align}
\end{subequations}
which places the poles at $\boldsymbol{p}_\mathrm{z1}$ and $\boldsymbol{p}_\mathrm{z2}$.
The gains can be selected based on desired continuous-time pole locations and pole matching as $\boldsymbol{p}_\mathrm{z}=\mathrm{e}^{\boldsymbol{p}_\mathrm{s}\Ts}$, where $\boldsymbol{p}_\mathrm{z}$ is the discrete-time pole and $\boldsymbol{p}_\mathrm{s}$ is the corresponding continuous-time pole.

\subsection{Sequence Separation}
The first-order sequence-separation filter \eqref{eq:sequence_filter} can be discretized as
\begin{subequations}
\begin{align}
    \ifilt(k) &= \af\ifilt(k-1) + \frac{1 - \af}{1 + \jj}\ic(k)\\
    \af &= \frac{\jj2\boldsymbol{\phi}_\mathrm{g}^\mi}{1 + \jj - (1-\jj)\boldsymbol{\phi}_\mathrm{g}^\mi}.
\end{align}
\end{subequations}
The discrete-time filter is designed such that in the periodic steady state the filter has the gains $1/(1+\jj)$ and $1/(1-\jj)$ for the positive and negative sequences, respectively. This makes the sequence separation exact, assuming that $\omegac=\omegag$.

\section{Effect of Inductance Estimation Error} \label{app:inductance_error}
A nonzero inductance estimation error $\tL = L - \hL$ is considered. Other assumptions are the same as in Section~\ref{chap:lin}, i.e., $\omegac=\omegag$ and $\ucref=\uc$, and the converter operates in grid-forming mode within its current limits.

\subsection{Disturbance Observer}

Using the system model \eqref{eq:model} and the disturbance observer \eqref{eq:obs_cont}, the estimation-error dynamics of the positive- and negative-sequence grid voltage become
\begin{subequations}\label{eq:estimation_error_coupled}
\begin{align}
    \frac{\D \Delta\tildeugp}{\D t} &= -\frac{\hL}{L}\boldsymbol{k}_\mathrm{o1}\left(\Delta\tildeugp + \Delta\tildeugn\right) - \frac{\tL}{L}\boldsymbol{k}_\mathrm{o1}\Delta\ec\nonumber\\
    &\quad - \jj\omegag\frac{\tL \hL}{L}\boldsymbol{k}_\mathrm{o1}\left(\Delta\icp - \Delta\icn\right)\label{eq:ugp_error_dyn}\\
    \frac{\D \Delta\tildeugn}{\D t} &= -\frac{\hL}{L}\boldsymbol{k}_\mathrm{o2} \left(\Delta\tildeugp + \Delta\tildeugn\right) - \jj2\omegag\Delta\tildeugn\nonumber\\
    &\quad - \frac{\tL}{L}\boldsymbol{k}_\mathrm{o2}\Delta\ec - \jj\omegag\frac{\tL \hL}{L}\boldsymbol{k}_\mathrm{o2}\left(\Delta\icp - \Delta\icn\right).
\end{align}
\end{subequations}
The inductance estimation error introduces a coupling from the control law to the estimation-error dynamics through $\ec$.

\subsection{Closed-Loop Control Dynamics}
The closed-loop system is linearized around an operating point in the periodic steady state. The angle of the coordinate system is aligned with the operating-point converter voltage, i.e., $\vco=\vcpmago$. From the output dynamics \eqref{eq:output_dyn} [which originate from \eqref{eq:model} and \eqref{eq:augmented_model}] and the control law \eqref{eq:gfm_control_law}, the closed-loop dynamics follow:
\begin{subequations} \label{eq:dyn_ctrl_coupled}
\begin{align}
    \frac{\D\Delta\pgp}{\D t} &= \alphap(\Delta\pgref-\Delta\pgp) + \betapv(\Delta\vcref-\Delta\vcpmag) \nonumber\\
    &\quad - \RE\{\betapn'\Delta\vcn\} - \gamma_\mathrm{pp}\Delta\pgp - \gamma_\mathrm{pv}\Delta\vcpmag\nonumber\\
    &\quad + \frac{1}{L}\RE\left\{\frac{(\ugpo)^*}{1+\jj}\ru_\mathrm{o}\right\}\\
    \frac{\D\Delta\vcpmag}{\D t} &= \alphav(\Delta\vcref-\Delta\vcpmag) + \betavp(\Delta\pgref-\Delta\pgp)\nonumber\\
    &\quad - \RE\{\betavn'\Delta\vcn\} - \gamma_\mathrm{vp}\Delta\pgp - \gamma_\mathrm{vv}\Delta\vcpmag\nonumber\\
    &\quad + \omegag\RE\left\{\frac{\ru_\mathrm{o}}{1-\jj}\right\}\\
    \frac{\D\Delta\vcn}{\D t} &= -\alphan'\Delta\vcn + \betanp(\Delta\pgref-\Delta\pgp)\nonumber\\
    &\quad + \betanv(\Delta\vcref-\Delta\vcpmag) - \boldsymbol{\gamma}_\mathrm{np}\Delta\pgp\nonumber\\
    &\quad  - \boldsymbol{\gamma}_\mathrm{nv}\Delta\vcpmag + \frac{\omegag}{1+\jj}\ru_\mathrm{o}
\end{align}
where the voltage-coupling term from the observer to the control law is
\begin{align}
    \ru_\mathrm{o} =& -\Delta\tildeugp + \kpbo\RE\{\Delta\tildeugp(\icpo)^*\} + \kvbo\RE\{\Delta\tildeugp\}\nonumber\\
    &\quad + (\rkno-1)\Delta\tildeugn.
\end{align}
The additional coefficients originating from the inductance estimation error are
\begin{gather}
    \gamma_\mathrm{pp} = \frac{\tL}{L^2}\RE\left\{\frac{(\ugpo)^*\boldsymbol{a}_\mathrm{p}}{1+\jj}\right\} \qquad
    \gamma_\mathrm{pv} = \frac{\tL}{L^2}\RE\left\{\frac{(\ugpo)^*\boldsymbol{a}_\mathrm{v}}{1+\jj}\right\}\nonumber\\
    \gamma_\mathrm{vp} = \frac{\omegag\tL}{L}\RE\left\{\frac{\boldsymbol{a}_\mathrm{p}}{1-\jj}\right\} \qquad
    \gamma_\mathrm{vv} = \frac{\omegag\tL}{L}\RE\left\{\frac{\boldsymbol{a}_\mathrm{v}}{1-\jj}\right\}\nonumber\\
    \boldsymbol{\gamma}_\mathrm{np} = \frac{\omegag\tL}{L}\,\frac{\boldsymbol{a}_\mathrm{p}}{1+\jj} \qquad
    \boldsymbol{\gamma}_\mathrm{nv} = \frac{\omegag\tL}{L}\,\frac{\boldsymbol{a}_\mathrm{v}}{1+\jj}
\end{gather}
where
\begin{gather}
    \boldsymbol{a}_\mathrm{p} = \jj\frac{\omegag L}{\rugdo} - \kpbo\frac{\rugdo-\rvcpo}{\rugdo}\nonumber\\
    \boldsymbol{a}_\mathrm{v} = 1 - \jj\tan\delta_0 - \kpbo\frac{\pgpo}{\rugdo} - \kvbo\nonumber\\
    \rugdo=\RE\{\ugpo\} \qquad \tan\delta_0=\frac{\omegag L}{\rvcpo\rugdo}\pgpo
\end{gather}
and $\delta_0$ is the operating-point load angle. The expression for $\tan\delta_0$ follows from the load-angle equation \eqref{eq:load_angle} and $\rugdo=|\ugpo|\cos\delta_0$.
The negative-sequence coefficients depending on the inductance estimation error are
\begin{gather}
    \alphan' = \omegag\frac{\rknoe + \jj(4-\rknoe)}{2}\qquad
    \betapn' = \frac{\rknoe(\ugpo)^*}{(1+\jj)L}\nonumber\\
    \betavn' = \frac{\omegag\rknoe}{1-\jj}
\end{gather}
where
\begin{align}
    \rknoe = \frac{\hL}{L}\rkno + \frac{\tL}{L}
\end{align}
is the effective negative-sequence gain. For $\hL=L$, \eqref{eq:estimation_error_coupled} and \eqref{eq:dyn_ctrl_coupled} reduce to \eqref{eq:estimation_error_dyn} and \eqref{eq:dyn_linear}.
\end{subequations}

\IEEEtriggeratref{29}


\end{document}